\documentclass[aip,jap,preprint,amsmath,amssymb,floatfix]{revtex4-2}

\usepackage{graphicx}
\usepackage{bm}
\usepackage{hyperref}

\begin{document}

\title{Hysteretic phononic band structures arising from martensitic phase transformations}

\author{R. Esquivel-Sirvent}
\affiliation{Instituto de F\'isica, Universidad Nacional Aut\'onoma de
M\'exico, Apdo.~Postal 20-364 CDMX 01000, Mexico}
\email{raul@fisica.unam.mx}

\author{B. Manzanares-Mart\'inez}
\author{J. Manzanares-Mart\'inez}
\affiliation{Departamento de Investigaci\'on en F\'isica de la
Universidad de Sonora, Apartado Postal 5-088, Hermosillo, Sonora 83190,
Mexico}

\date{\today}

\begin{abstract}
Thermal tuning of phononic crystals typically treats each constituent's elastic modulus as a single-valued function of temperature. Here we show that when one constituent undergoes a first-order martensitic transformation (NiTiCu), paired with a Parylene C spacer in a thin one-dimensional composite rod, the intrinsic thermal hysteresis of the transformation is transferred to the collective acoustic response, making it depend on thermal history. Transfer-matrix calculations performed separately along the heating and cooling branches reveal that the same temperature within the transformation interval can correspond to two distinct Bloch band structures and two distinct transmission spectra. As temperature is cycled, stop-band edges trace closed loops in the temperature-frequency plane, producing hysteretic phononic band structures. At a fixed probe frequency, the rod can switch between transmitting and strongly attenuating states depending solely on whether the sample was last heated or cooled, with transmission contrasts exceeding 50 dB in a six-cell structure, thereby realizing a form of acoustic memory. The filling fraction of the active segment provides an independent geometric control parameter that modifies the width of the hysteresis loops and the positions of gap closures. These predictions can be tested directly using immersion-ultrasonic measurements on NiTiCu/Parylene C composite rods.
\end{abstract}

\maketitle

\section{Introduction}
\label{sec:intro}

Phononic crystals and acoustic metamaterials constitute a widely
studied class of artificial media for controlling elastic-wave
propagation through engineered periodicity and material
contrast~\cite{Muhammad2022,Chen2020review,Vasileiadis2021,Pennec2016}.
Early studies established the basic concepts of acoustic band
gaps~\cite{EsquivelCocoletzi1994,Kushwaha1993,Kushwaha1994}, dispersion
engineering in solid structures~\cite{Morales2002,DiazAnda2005}, and
wave localization in multilayered and lattice-based structures.

Beyond longitudinal acoustic waves in fluids or effectively
one-dimensional media, significant advances have been achieved in the
control of elastic bulk and surface waves in periodic and layered
solids~\cite{Morales2002,DiazAnda2005,Achenbach2012}. Periodic elastic
structures support multiple polarizations and mode
families~\cite{Manzanares2010,ManzanaresRamos2003,CastroArce2010},
leading to dispersion relations that are considerably richer than
their acoustic
counterparts~\cite{Podlipenets1987,ManzanaresBrewster2000,ManzanaresSagittal2007}.
In plates and layered substrates, for example, the propagation of Lamb
waves has been extensively studied in phononic crystal
configurations~\cite{Yuan2020,Yao2011}. Periodic patterning or layering
can open band gaps for specific Lamb-wave branches, induce strong mode
repulsion and hybridization, and generate flat bands associated with
slow-wave or resonant behavior~\cite{Kherraz2019}. These effects have
been exploited for wave guiding, vibration isolation,
frequency-selective filtering, and enhanced wave--matter
interactions~\cite{Liu2013}.

A central issue in phononic-crystal research is the search for
external control parameters that allow continuous modulation of
acoustic properties without mechanical
reassembly~\cite{ChenReview2018}, that can facilitate band-engineering.
Approaches based on mechanical deformation~\cite{Hu2025}, electric or
magnetic fields~\cite{ChenZ2016,Gardiner2024}, and active elements have
been explored. Thermal control offers an appealing alternative:
temperature can be applied locally or globally, is compatible with
micro- and mesoscale architectures, and directly modifies intrinsic
material properties such as elastic moduli and sound
velocities~\cite{Wei2020,Guo2024,Jin2025}. In particular, the use of
shape-memory alloy inserts to control wave propagation in periodic
rods was pioneered by Ruzzene and Baz two decades
ago~\cite{RuzzeneBaz2000a,RuzzeneBaz2000b,ChenRuzzeneBaz2000}, and the
idea has since been extended to two-dimensional crystals, plates, and
defective phononic structures~\cite{Yao2011,ChenZJ2015,Jo2025}.
However, in many thermally tunable or phase-change-based phononic
systems, the elastic modulus of the active material is treated as a
single-valued function of temperature: the band structure is computed
at each temperature under the assumption that the elastic constants
are uniquely determined by the instantaneous thermal state.

Real first-order structural transitions, however, are not
single-valued. The martensitic transformation in NiTi-based alloys
proceeds along distinct trajectories upon heating and cooling, with
the austenitic and martensitic finish temperatures separated by a
finite hysteresis width that is intrinsic to the cooperative lattice
rearrangement~\cite{Gisser1994,Huibin1991,Nespoli2011}. Within the
transformation interval, the same external temperature corresponds to
two different phase fractions, and consequently to two different
values of the elastic modulus, depending on whether the system has
been heated from below or cooled from above. This thermal hysteresis
has been characterized by various experimental techniques, including
the recent acoustic measurement of $E(T)$ by Rozzi
\textit{et al.}~\cite{Rozzi2025}, which provides the quantitative
input for the present model.

Shape-memory alloys provide a particularly attractive material
platform. NiTi-based alloys exhibit reversible martensite-austenite
phase transformations over experimentally accessible temperature
ranges, accompanied by pronounced and continuous changes in elastic
stiffness~\cite{Uddin2025,Sewak2019,Ledbetter1973}. These changes in
density and elastic constants can be translated directly into tunable
acoustic phase velocities and impedances, enabling temperature to act
as an effective control knob for elastic-wave propagation. Previous
studies have considered thermal tuning of phononic band gaps in
nitinol-based periodic systems~\cite{Yao2011,ChenZJ2015}. Here we
focus on a distinct regime in which the active material itself
possesses intrinsic thermal hysteresis, so that the acoustic response
depends not only on the temperature but also on the thermal path. A
related but different question was examined by Jang
\textit{et al.}~\cite{Jang2009}, who used a shape-memory polymer to
lock a two-dimensional pattern into one of two states and observed
switching between two phononic spectra (see
also~\cite{BertoldiBoyce2008}). That mechanism is bistable in geometry
rather than in material parameters; the case of a continuous,
temperature-driven hysteresis in the elastic constants of a
constituent has not, to our knowledge, been analyzed for periodic
elastic systems.

In this work, we investigate thin one-dimensional composite rods
composed of alternating segments of a temperature-sensitive
shape-memory alloy (NiTiCu) and a polymer spacer (Parylene~C). In the
thin-rod regime, the longitudinal phase velocity reduces to
$c=\sqrt{E/\rho}$, which coincides with the regime in which $E(T)$ has
been measured for NiTiCu~\cite{Rozzi2025}; the experimental data
therefore enter the model directly without reinterpretation of the
elastic constants. The polymer segments provide acoustic impedance
contrast while also acting as partial thermal barriers between
adjacent metallic segments. We show that the martensitic
transformation of NiTiCu induces substantial modifications of the
phononic band structure, including thermally controlled shifts of stop
bands and, most notably, hysteretic band structures in which the same
temperature can correspond to different transmission spectra depending
on whether the system is heated or cooled.

\section{Thermal control of acoustic impedance in NiTiCu memory alloys}
\label{sec:material}

The temperature-dependent acoustic properties of NiTiCu---more
specifically, the ternary composition Ni$_{40}$Ti$_{50}$Cu$_{10}$ that
we adopt throughout this work, following the experimental
characterization of Rozzi \textit{et al.}~\cite{Rozzi2025}---are rooted
in the martensitic phase transformation that characterizes shape-memory
alloys. As temperature increases, NiTiCu undergoes a diffusionless
solid--solid transformation from a low-symmetry martensitic phase to a
high-symmetry austenitic
phase~\cite{Sewak2019,Gisser1994,Nishiyama2012}. This transformation is
first order in nature and therefore exhibits coexistence of phases and
thermal hysteresis; it involves a cooperative lattice rearrangement,
and is accompanied by pronounced changes in elastic
stiffness~\cite{Ledbetter1973,Ba2025}. Importantly, the transformation
does not occur at a single temperature but extends over a finite
interval, within which both phases coexist and the relative phase
fraction evolves continuously. Thus, the Young modulus can be written
as a weighted average~\cite{Ledbetter1973,Zotov2014,Rozzi2025}
\begin{equation}
E(T) = \zeta(T)\,E_a + \bigl[1 - \zeta(T)\bigr]\,E_m,
\label{eq:Eofzeta}
\end{equation}
where $E_m = 26$~GPa is the Young modulus in the martensitic state and
$E_a = 69$~GPa in the austenitic phase, both extracted from the
acoustic measurements of Rozzi \textit{et al.}~\cite{Rozzi2025}. The
weight function $\zeta(T)$ is determined as
\begin{equation}
\zeta(T) = \frac{1}{1 + e^{\,k(T-T_c)}},
\label{eq:zeta}
\end{equation}
where $k = -1.4~^\circ\text{C}^{-1}$ and $T_c = 38~^\circ\text{C}$,
obtained from the same experimental
measurements~\cite{Rozzi2025}.

Those measurements correspond to the cooling branch of the
transformation. Since NiTiCu exhibits hysteresis, the heating branch
generally follows a different
trajectory~\cite{Gisser1994,Huibin1991,Nespoli2011}. Since the heating
branch was not reported in Ref.~\cite{Rozzi2025}, the hysteresis width
cannot be uniquely determined from available data. The heating branch
was therefore constructed based on the expected qualitative behavior
of the hysteresis loop by shifting the temperature of the transition
midpoint to $T_c = 48~^\circ\text{C}$, while preserving the same
functional form of the process. The resulting hysteresis width
$\Delta T \approx 10~^\circ\text{C}$ is consistent with values
commonly reported for NiTiCu compositions of this
family~\cite{Otsuka2005,Nespoli2011,Huibin1991}, and captures the
essential path dependence of the transformation.

The density varies weakly with temperature, since martensitic
transformations in NiTi-based alloys are diffusionless and largely
volume-conserving~\cite{Ledbetter1973,Bhattacharya2003}. In the
martensitic phase the density is $\rho_m = 6170~\text{kg/m}^3$, while
in the austenitic phase it is
$\rho_a = 6460~\text{kg/m}^3$~\cite{Otsuka2005}. The temperature
dependence of the density can be described using the same weight
function $\zeta(T)$ introduced in Eq.~\eqref{eq:zeta}, such that
\begin{equation}
\rho(T) = \zeta(T)\,\rho_a + \bigl[1 - \zeta(T)\bigr]\,\rho_m.
\label{eq:rhoofT}
\end{equation}
The contribution of ordinary thermal expansion to $\rho(T)$ is of
order $10^{-5}\,^\circ\text{C}^{-1}$ and is therefore negligible
compared to the change associated with the phase transformation; we
omit it in what follows.

More comprehensive thermomechanical descriptions, such as the Brinson
model~\cite{Brinson1993}, explicitly account for stress--temperature
coupling and the evolution of internal variables. However, although
acoustic waves involve dynamic stress fields, their amplitudes are
typically insufficient to modify the phase fraction. Consequently, the
material properties can be treated as functions of temperature alone.

The behavior of the Young modulus $E(T)$ and density $\rho(T)$ has
direct and important consequences. In the thin-rod regime, where the
lateral cross section of the rod is much smaller than the relevant
acoustic wavelengths, the longitudinal phase velocity reduces to
\begin{equation}
c(T) = \sqrt{\frac{E(T)}{\rho(T)}}.
\label{eq:cofT}
\end{equation}
This is the same regime in which the elastic modulus of NiTiCu was
measured by Rozzi \textit{et al.}~\cite{Rozzi2025}, so the values of
$E_m$ and $E_a$ used in Eq.~\eqref{eq:Eofzeta} can be applied directly
in Eq.~\eqref{eq:cofT} without reinterpretation of the elastic
constants. The corresponding specific acoustic impedance is
\begin{equation}
Z(T) = \rho(T)\,c(T) = \sqrt{E(T)\,\rho(T)}.
\label{eq:ZofT}
\end{equation}

The asymptotic values of $E$, $\rho$, $c$, and $Z$ in the two pure
phases are summarized in Table~\ref{tab:phaseparams}. The phase
velocity changes by a factor of $\sim\!1.6$ between martensite and
austenite, dominated by the variation in the elastic modulus; the
change in density gives a much smaller correction. The specific
acoustic impedance varies by a comparable factor between the two
phases.

\begin{table}[t]
\caption{Asymptotic acoustic parameters of the NiTiCu segments in the
two pure phases. Young modulus and density are taken from
Refs.~\cite{Rozzi2025,Otsuka2005}. Phase velocity and specific acoustic
impedance follow from Eqs.~\eqref{eq:cofT}--\eqref{eq:ZofT}.}
\label{tab:phaseparams}
\begin{ruledtabular}
\begin{tabular}{lcc}
 & Martensite ($\zeta=0$) & Austenite ($\zeta=1$) \\
\hline
$E$ (GPa)            & 26    & 69    \\
$\rho$ (kg/m$^3$)    & 6170  & 6460  \\
$c$ (m/s)            & 2053  & 3268  \\
$Z$ (MRayl)          & 12.67 & 21.11 \\
\end{tabular}
\end{ruledtabular}
\end{table}

\begin{figure}[t]
\centering
\includegraphics[width=0.95\columnwidth]{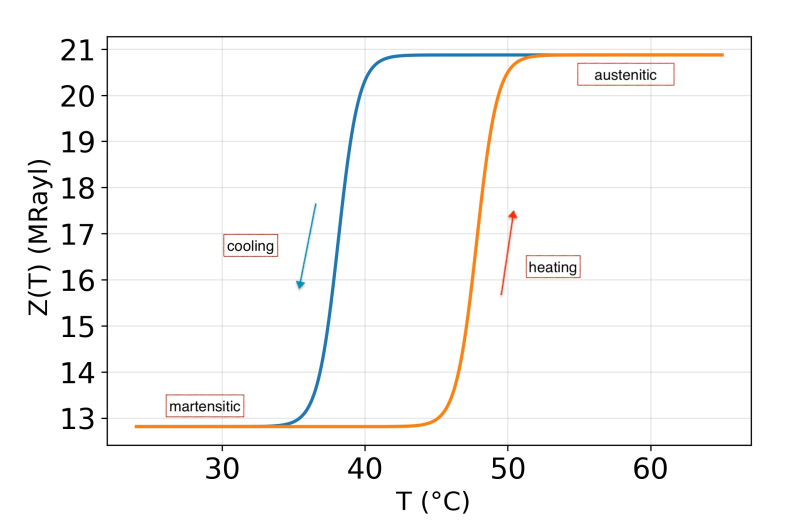}
\caption{Temperature dependence of the specific acoustic impedance of
NiTiCu during cooling (blue) and heating (orange), illustrating the
intrinsic thermal hysteresis of the alloy. The asymptotic values
($Z_m \approx 12.7$~MRayl in the martensitic plateau and
$Z_a \approx 21.1$~MRayl in the austenitic plateau) coincide with the
values reported in Ref.~\cite{Rozzi2025}, validating the calibration
of the material model. The horizontal separation between the two
branches is the hysteresis width $\Delta T \approx 10~^\circ$C.}
\label{fig:Zloop}
\end{figure}

Thus, the martensitic transformation translates directly into a
strong, continuous modulation of the acoustic impedance. In
Fig.~\ref{fig:Zloop}, we show the specific acoustic impedance of
NiTiCu as a function of temperature. The blue and orange curves
correspond to the cooling and heating branches, respectively,
illustrating the intrinsic thermal hysteresis of the alloy. The
asymptotic values $Z_m\approx 12.7\,\text{MRayl}$ (martensite) and
$Z_a\approx 21.1\,\text{MRayl}$ (austenite) coincide with those
reported in Ref.~\cite{Rozzi2025}, providing a direct validation of
the calibration of the material model. Within the transformation
interval, the same external temperature corresponds to two different
phase fractions and therefore to two different impedance values,
depending on whether the system has been heated from below or cooled
from above. This path-dependent acoustic impedance is the
material-level signature of the underlying first-order transition and
is the input that the periodic structure (Sec.~III) translates into a
path-dependent Bloch spectrum.

\section{Layered system}
\label{sec:system}

\begin{figure}[t]
\centering
\includegraphics[width=\columnwidth]{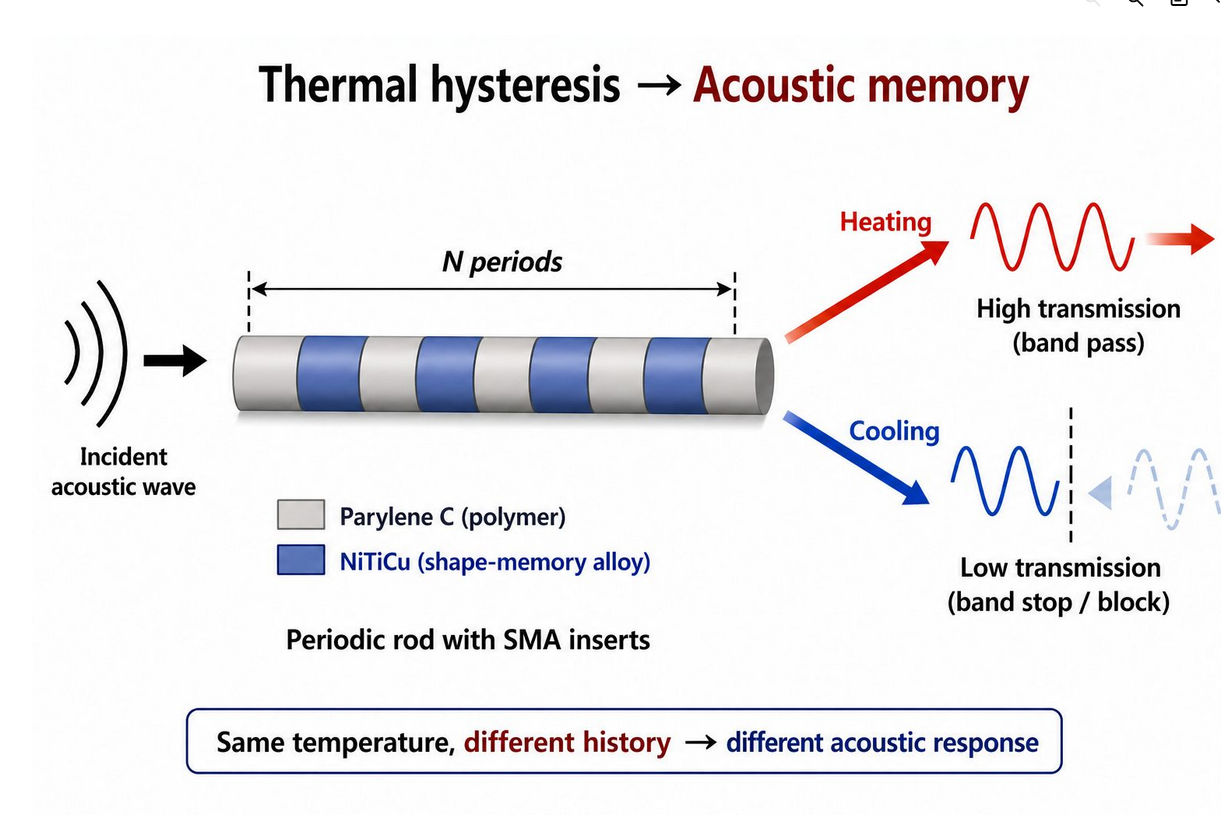}
\caption{Schematic of the composite phononic rod under study. A
one-dimensional periodic rod is formed by alternating cylindrical
segments of Parylene~C (gray) and NiTiCu (blue), with $N$ unit cells
of total period $d=d_{1}+d_{2}$. An incident longitudinal acoustic
wave probes the structure. Within the transformation interval of
NiTiCu, the same external temperature can correspond to a
high-transmission (band-pass) state if the system is being heated, or
to a low-transmission (band-stop) state if the system is being cooled,
illustrating the central effect analyzed in this work: identical
temperatures, different thermal histories, distinct acoustic
responses.}
\label{fig:rod}
\end{figure}

We consider a one-dimensional phononic rod made of alternating
segments of a polymer (Parylene~C) and the NiTiCu shape-memory alloy,
as depicted schematically in Fig.~\ref{fig:rod}. The polymer segments
serve a dual purpose. Acoustically, they provide impedance contrast
with respect to the metallic alloy, thereby enabling Bragg scattering
and stop-band formation. Thermally, Parylene~C is a low-conductivity
polymer~\cite{Guermoudi2021} that partially isolates adjacent NiTiCu
segments, a desirable property for any future implementation in which
independent local heating of individual segments is contemplated.
Parylene~C is also well established in ultrasonic transducers and
acoustic devices~\cite{Levassort2005,Hadimioglu1990}; its density is
$\rho = 1280$~kg/m$^{3}$, the longitudinal speed of sound is
$c_L = 2135$~m/s, and the corresponding specific acoustic impedance is
$Z = 2.7$~MRayl~\cite{Sim2005}.

We work throughout in the thin-rod regime, in which the lateral cross
section of the rod is much smaller than the relevant acoustic
wavelength. In this limit, lateral strains are unconstrained and the
longitudinal phase velocity in each segment reduces to
$c=\sqrt{E/\rho}$, the same regime in which the elastic modulus of
NiTiCu was measured by Rozzi \textit{et al.}~\cite{Rozzi2025}. For
the frequency range explored in this work,
$\nu \in [1, 20]$~MHz, and acoustic velocities of order
$2$--$3$~km/s, the corresponding wavelengths are in the range
$\lambda \in [100, 3000]~\mu$m. The thin-rod approximation is
therefore valid as long as the rod diameter remains well below
$\sim 100~\mu$m, a constraint readily satisfied by commercially
available NiTi wires used in MEMS and medical applications.
Extensions to thicker rods would require including geometric
(Pochhammer--Chree) dispersion~\cite{Achenbach2012,Graff1991}, which would shift the absolute
positions of the gaps at high frequency but would not modify the
hysteretic nature of the phenomenon.

We consider two cases: an infinite periodic superlattice, which yields
the Bloch band structure of the ideal system, and a finite composite
rod, which provides the transmission response relevant to
experimental realizations. In both cases, the elementary unit cell
consists of one Parylene~C segment of thickness $d_{1}$ and one NiTiCu
segment of thickness $d_{2}$, so that the lattice period is
\begin{equation}
d = d_{1} + d_{2}.
\label{eq:period}
\end{equation}
It is convenient to define the filling fraction of the active NiTiCu
segment as
\begin{equation}
\phi = \frac{d_{2}}{d_{1} + d_{2}},
\label{eq:phi}
\end{equation}
with the complementary polymer fraction
$1 - \phi = d_{1}/(d_{1} + d_{2})$. The parameter $\phi$ provides a
direct geometrical control variable that modifies both the acoustic
path length and the effective impedance contrast within the unit cell. 

The transfer matrix relates the acoustic displacement and stress
fields across each segment. We adopt as state vector the pair
$\Psi(x) = (u(x), \sigma(x))^{T}$, where $u$ is the longitudinal
displacement and $\sigma$ is the axial stress. The choice of
$(u, \sigma)$ is convenient because both quantities are continuous at
any solid--solid interface in perfect contact, so that no separate
interface matrix is required. For a uniform segment $j$ of thickness
$d_{j}$, wave number $k_{j} = \omega/c_{j}$, and specific acoustic
impedance $Z_{j} = \rho_{j} c_{j}$, the propagator that maps the
state at the left edge to the state at the right edge
is~\cite{PerezAlvarez2004,Manzanares2012}
\begin{equation}
M_{j}(d_{j})
=
\begin{pmatrix}
\cos(k_{j}d_{j}) & \dfrac{\sin(k_{j}d_{j})}{\omega Z_{j}} \\[10pt]
-\,\omega Z_{j}\,\sin(k_{j}d_{j}) & \cos(k_{j}d_{j})
\end{pmatrix}.
\label{eq:Mj}
\end{equation}

For the bilayer unit cell composed of Parylene~C followed by NiTiCu,
the unit-cell transfer matrix is
\begin{equation}
M_{u} = M_{2}(d_{2})\,M_{1}(d_{1}),
\label{eq:Mu}
\end{equation}
where the order of multiplication follows the order of propagation
through the cell. For the infinite case we have the usual dispersion
relation
\begin{equation}
\cos(qd) = \tfrac{1}{2}\,\text{Tr}(M_{u}),
\label{eq:Bloch}
\end{equation}
where $q$ is the Bloch wave vector. Allowed propagating bands
correspond to $|\tfrac{1}{2}\text{Tr}(M_{u})| \le 1$, where $q$ is
real and Bloch waves propagate freely; forbidden gaps correspond to
$|\tfrac{1}{2}\text{Tr}(M_{u})| > 1$, where $q$ becomes complex and
waves decay exponentially. Explicit evaluation of the trace yields
the closed-form dispersion relation in terms of the filling fraction
$\phi$,
\begin{multline}
\cos(qd) = \cos\!\left[\frac{\omega(1-\phi)d}{c_{1}}\right]
\cos\!\left[\frac{\omega\phi d}{c_{2}}\right] \\[2pt]
- \frac{1}{2}\!\left(\frac{Z_{1}}{Z_{2}} + \frac{Z_{2}}{Z_{1}}\right)
\sin\!\left[\frac{\omega(1-\phi)d}{c_{1}}\right]
\sin\!\left[\frac{\omega\phi d}{c_{2}}\right],
\label{eq:cosFormula}
\end{multline}
where the subscripts 1 and 2 refer to the Parylene and NiTiCu
segments, respectively.  The dispersion relation is controlled by the impedance ratio Z2/Z1 and the phase advances $\omega(1-\phi)d/c_{1}$ and $\omega\phi d/c_{2}$.

For a finite periodic structure with $N$ unit cells, the total
transfer matrix is
\begin{equation}
M_{\text{tot}} = (M_{u})^{N}
\equiv
\begin{pmatrix} M_{11} & M_{12} \\ M_{21} & M_{22} \end{pmatrix},
\label{eq:Mtot}
\end{equation}
or, if additional capping segments are included, the corresponding
segment matrices must be multiplied on the left or right according
to their physical order. The transmission amplitude can be obtained
from the total transfer matrix by matching the acoustic fields to
the incident and transmitted media. If the incident and exit media
have impedances $Z_{L}$ and $Z_{R}$, respectively, then the
transmission amplitude is
\begin{equation}
t = \frac{2\,i\omega Z_{L}}
{\bigl(M_{11} + i\omega Z_{R}\,M_{12}\bigr)i\omega Z_{L}
+ M_{21} + i\omega Z_{R}\,M_{22}}\,.
\label{eq:tamp}
\end{equation}
The corresponding power transmission coefficient is
\begin{equation}
\tau = \frac{Z_{L}}{Z_{R}}\,|t|^{2},
\label{eq:tau}
\end{equation}
which reduces to $\tau = |t|^{2}$ for the symmetric case in which the
composite rod is immersed in water on both sides
($Z_{L} = Z_{R} = Z_{\text{water}}$).

The infinite and finite formulations are complementary: the former
provides the intrinsic Bloch band structure of the ideal crystal,
while the latter captures finite-size resonances and the
experimentally observable transmission spectra.

\begin{figure}[t]
\centering
\includegraphics[width=\columnwidth]{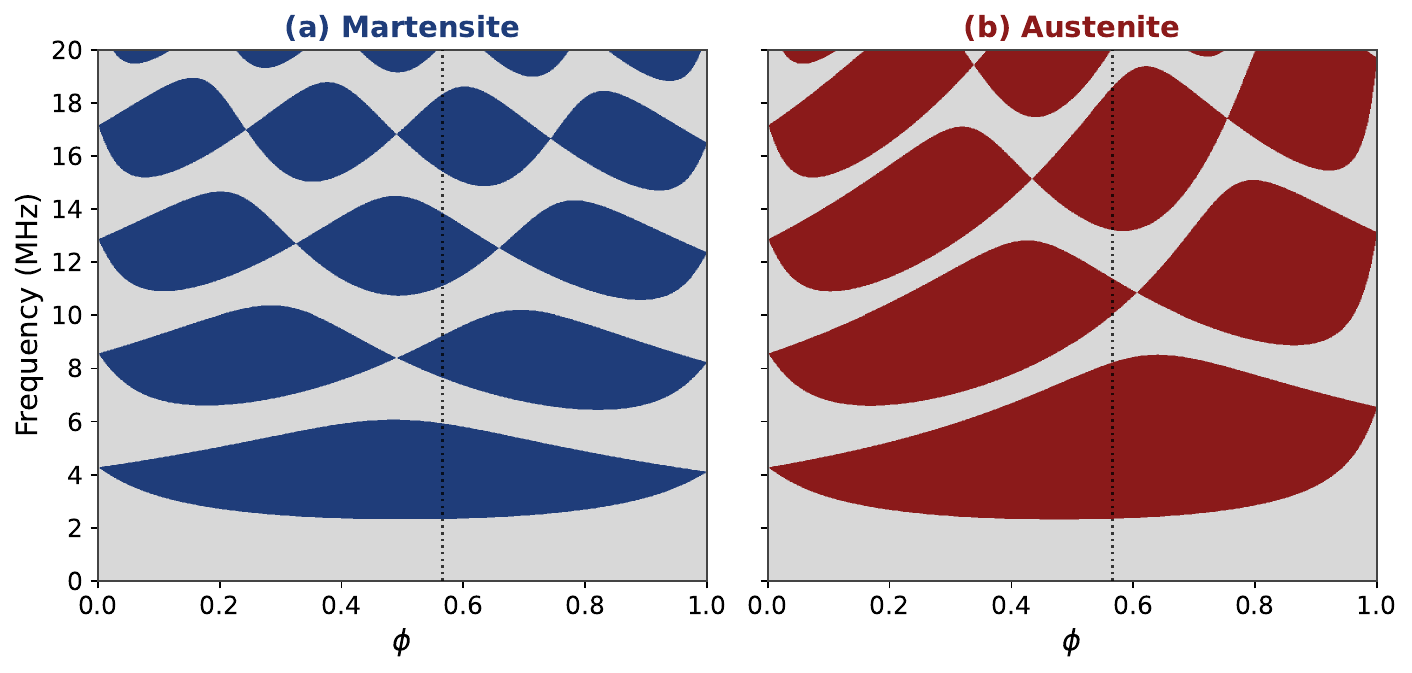}
\caption{Band-gap maps of the infinite periodic rod in the
$(\phi, \nu)$ plane, computed from the cosine
formula~\eqref{eq:cosFormula} for the two pure phases of NiTiCu.
(a) Martensite ($c_{2} = c_{m} = 2053$~m/s,
$Z_{2} = Z_{m} = 12.7$~MRayl). (b) Austenite
($c_{2} = c_{a} = 3268$~m/s, $Z_{2} = Z_{a} = 21.1$~MRayl). Colored
regions denote forbidden band gaps; gray regions denote propagating
bands. Both panels share the same Parylene~C segment as material 1
and the same period $d$.}
\label{fig:bandgapPhi}
\end{figure}

Before turning to the thermal hysteresis, it is instructive to
examine the band structure of the infinite system in the two pure
phases of NiTiCu, considered separately, as a function of the
filling fraction $\phi$. Equation~\eqref{eq:cosFormula} can be
inverted directly: for given material parameters $(c_{1}, Z_{1})$
and $(c_{2}, Z_{2})$, the gaps correspond to the regions of the
$(\phi, \omega)$ plane where the right-hand side exceeds unity in
absolute value. Figure~\ref{fig:bandgapPhi} shows the resulting
maps for martensite (panel a, $c_{2} = c_{m}$, $Z_{2} = Z_{m}$) and
austenite (panel b, $c_{2} = c_{a}$, $Z_{2} = Z_{a}$), evaluated
with the asymptotic values of Table~\ref{tab:phaseparams}. Several
qualitative features are apparent. First, both phases display the
standard pattern of one-dimensional phononic crystals: each gap opens
at intermediate filling fractions, reaches a maximum width, and
closes at specific values of $\phi$ determined by the Bragg
condition. Second, in the frequency range $0$--$20$~MHz the
austenitic phase exhibits one more gap than the martensitic phase
(five robust gaps versus four), and the austenitic gaps are
systematically wider, reflecting the stronger impedance contrast in
the austenitic phase. Third, the gap closures occur
at similar values of $\phi$ in both phases, since these contact
points are essentially geometric and depend on the period of the unit
cell rather than on the specific impedance values. The figure
therefore identifies the two asymptotic spectra between which the
phononic response of the rod will evolve when the temperature is
varied, the subject of the next section.

\section{Thermal tuning of the infinite periodic structure}
\label{sec:infinite}

We begin by analyzing the idealized infinite composite rod, which
provides the intrinsic Bloch band structure of the periodic crystal
without finite-size boundary effects. In this limit, the dispersion
relation~\eqref{eq:cosFormula} is governed solely by the unit-cell
geometry and the temperature-dependent elastic properties of the
NiTiCu segments. The infinite-system description therefore offers
the clearest framework for identifying propagating bands, stop gaps,
and their evolution through the martensitic transformation.

\begin{figure*}[t]
\centering
\includegraphics[width=0.95\textwidth]{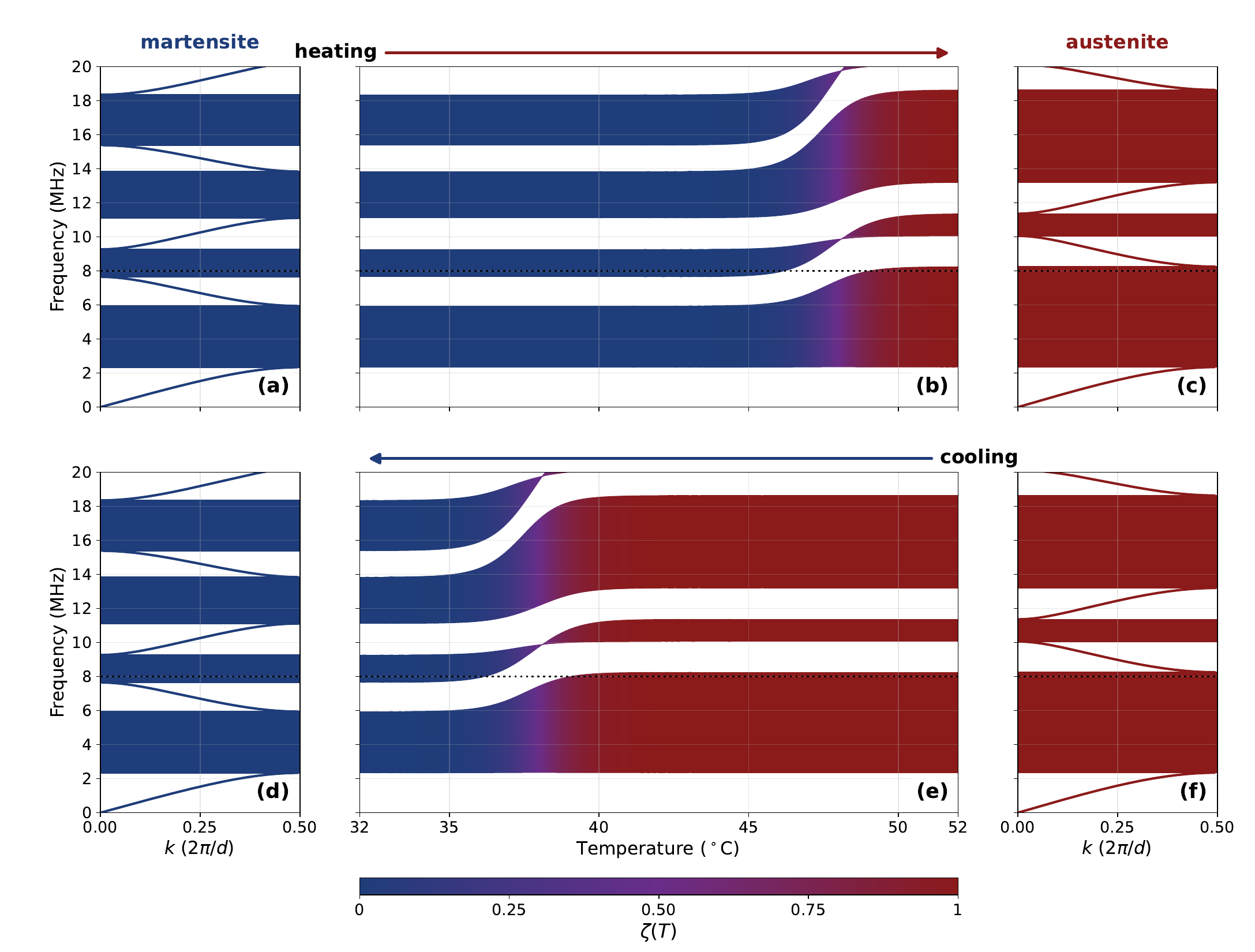}
\caption{Thermal evolution of the Bloch band structure of the
infinite periodic rod for the filling fraction $\phi = 0.566$
($d_{1} = 108~\mu$m, $d_{2} = 141~\mu$m). Panels (a)--(c) correspond
to the heating branch from martensite to austenite, while panels
(d)--(f) show the reverse cooling cycle. The central
temperature--frequency maps in panels (b) and (e) display the gaps
in colors that interpolate between blue (martensite) and red
(austenite) according to the local austenite fraction $\zeta(T)$.
Because the NiTiCu elastic properties follow different thermal paths
upon heating and cooling, the phononic band structure exhibits
pronounced hysteresis: the same temperature within the transformation
interval corresponds to different stop-gap positions depending on
the thermal branch.}
\label{fig:Tfmaps}
\end{figure*}

The evolution of the phononic spectrum is summarized in
Fig.~\ref{fig:Tfmaps} for a system with $d_{1} = 108~\mu$m
(Parylene) and $d_{2} = 141~\mu$m (NiTiCu), corresponding to the
filling fraction $\phi = 0.566$ examined in the asymptotic
$\phi$--$\nu$ maps of Fig.~\ref{fig:bandgapPhi}. The heating branch
is shown in the upper row, while the cooling branch is displayed in
the lower row, allowing direct comparison by vertical alignment. In
each row, the central temperature--frequency map is flanked by the
limiting band structures of pure martensite (left) and pure austenite
(right), so that the displacement of each gap across the
transformation can be followed continuously while retaining its
asymptotic endpoints. The lateral panels at $\phi = 0.566$ correspond
to the vertical slice of the asymptotic maps of
Fig.~\ref{fig:bandgapPhi} at this filling fraction, with each gap
appearing in the same absolute frequency window in both
representations.

As the phase fraction $\zeta(T)$ varies between zero and unity, the
stop gaps migrate smoothly between their martensitic and austenitic
positions. In addition, the overall gap-bearing spectral region
broadens in the austenitic phase, reflecting the larger elastic
stiffness and modified impedance contrast of the transformed NiTiCu
segments.

The central physical feature of Fig.~\ref{fig:Tfmaps} is the
horizontal offset between the heating and cooling branches. In the
present model, the heating transition is centered near
$T_{c}^{\,\text{heat}} = 48~^\circ$C, whereas the reverse cooling
branch is centered near $T_{c}^{\,\text{cool}} = 38~^\circ$C,
producing an effective hysteresis width
$\Delta T \approx 10~^\circ$C. Consequently, within the transformation
interval the same external temperature corresponds to different
phase fractions, and therefore to distinct phononic band structures
depending on thermal history.

This path dependence has an immediate operational consequence at
fixed frequency. Taking $\nu = 8$~MHz as a representative example,
the first stop gap of the martensitic crystal terminates below this
value, whereas in the austenitic phase the corresponding gap extends
above it. Thus, the low-temperature state is transmitting at
$8$~MHz, while the high-temperature state is strongly attenuating.
Because the heating and cooling branches cross this threshold at
different temperatures, an intermediate state such as
$T = 42~^\circ$C may correspond to either transmission or blocking
depending solely on the previous thermal path.

The periodic crystal therefore exhibits a thermally path-dependent
Bloch spectrum: the same sample, at the same temperature, can occupy
two distinct spectral states selected exclusively by thermal history.
This behavior is the direct manifestation of the material-level
hysteresis loop of Sec.~\ref{sec:material} in the collective
acoustic response of the periodic structure.

The infinite-periodic analysis therefore establishes the fundamental
mechanism of the present system: a reversible structural phase
transition in one constituent segment produces controllable and
hysteretic modifications of the Bloch spectrum. In the next section,
we show how these intrinsic changes manifest themselves in finite
composite rods through measurable transmission spectra.

\section{Finite composite rods and transmission spectra}
\label{sec:finite}

The infinite-periodic analysis of Sec.~\ref{sec:infinite} establishes
the intrinsic Bloch spectrum of the system, but any experimental
realization involves a finite stack of unit cells terminated by an
external medium. We now examine how the hysteresis of the underlying
material translates into the transmission spectrum of a finite
composite rod, the quantity directly accessible to immersion
ultrasonic measurements.

We consider a rod of $N = 6$ unit cells, with the same geometry used
in Sec.~\ref{sec:infinite}, $d_{1} = 108~\mu$m (Parylene) and
$d_{2} = 141~\mu$m (NiTiCu), so that the total length of the
composite region is $Nd \simeq 1.5$~mm. The rod is immersed in water
on both sides, $Z_{L} = Z_{R} = Z_{\text{water}} = 1.5$~MRayl, a
configuration relevant to standard ultrasonic
characterization~\cite{Levassort2005,Hadimioglu1990}. The power
transmission coefficient $\tau(\omega)$ is computed from the total
transfer matrix $M_{\text{tot}} = (M_{u})^{N}$ via
Eqs.~\eqref{eq:tamp}--\eqref{eq:tau}.

To probe the path-dependent character of the response, we evaluate
$\tau(\omega)$ at the two midpoints of the hysteresis loop,
$T = 38~^\circ$C (the cooling midpoint) and $T = 48~^\circ$C (the
heating midpoint). At each temperature, the two thermal branches
correspond to different austenite fractions:
\begin{itemize}
\item At $T = 38~^\circ$C: heating branch
$\zeta = 0$ (pure martensite); cooling branch $\zeta = 0.5$
(50/50 mixed phase).
\item At $T = 48~^\circ$C: heating branch $\zeta = 0.5$ (50/50
mixed phase); cooling branch $\zeta = 1$ (pure austenite).
\end{itemize}
The two midpoints are symmetric counterparts of the same loop:
each one places one branch in a pure phase and the other in the
fully mixed configuration, allowing the consequences of the thermal
path to be displayed at both extremes of the transformation
interval.

\begin{figure*}[t]
\centering
\includegraphics[width=0.95\textwidth]{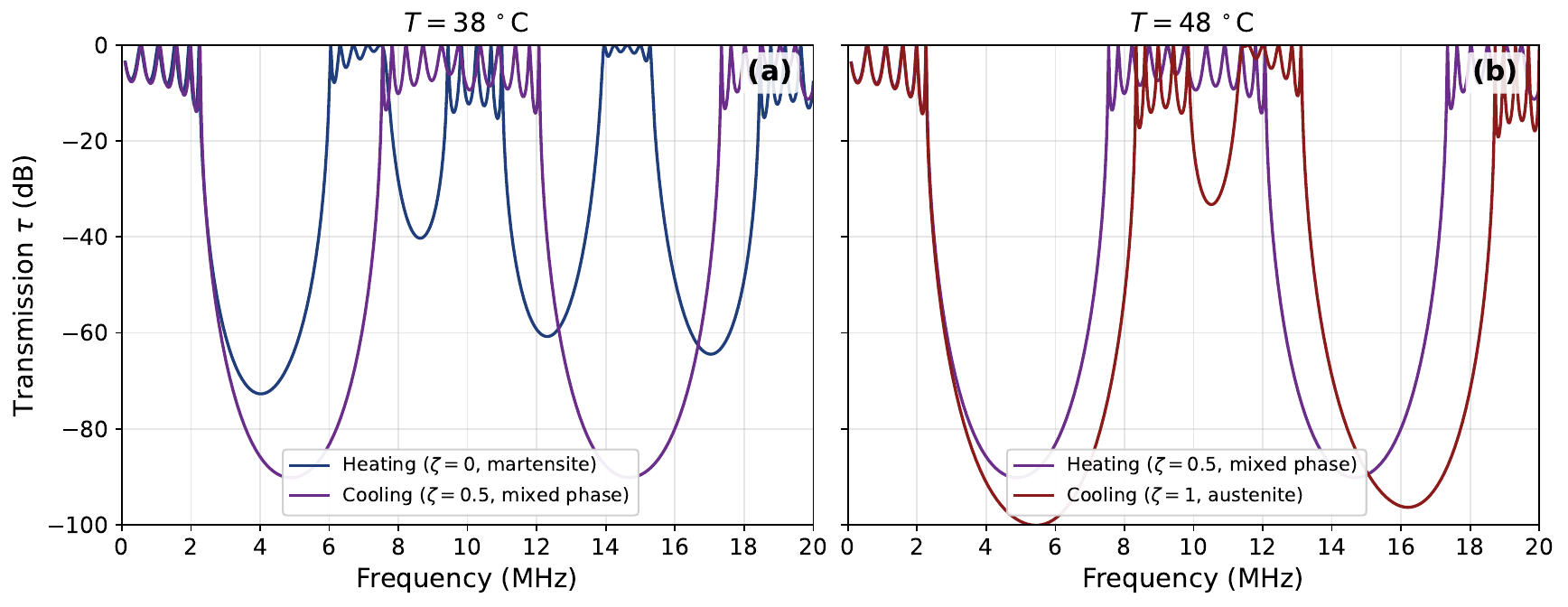}
\caption{Transmission spectra $\tau(\nu)$ of a finite composite rod
of $N = 6$ unit cells immersed in water, evaluated at the two
midpoints of the hysteresis loop. (a) $T = 38~^\circ$C: the heating
branch (blue, $\zeta = 0$) corresponds to pure martensite, while
the cooling branch (purple, $\zeta = 0.5$) corresponds to a
50/50 mixed phase. (b) $T = 48~^\circ$C: the heating branch
(purple, $\zeta = 0.5$) is the same mixed phase as in panel (a),
while the cooling branch (red, $\zeta = 1$) is pure austenite. At
each temperature, the two branches yield distinctly different
transmission spectra; this is the direct experimental manifestation
of the underlying material hysteresis.}
\label{fig:transmission}
\end{figure*}

The transmission spectra of Fig.~\ref{fig:transmission} exhibit the
characteristic features of a finite phononic crystal: well-defined
pass bands close to $\tau \simeq 0~$dB, separated by stop bands in
which $\tau$ drops by several tens of decibels. The fast oscillations
within each pass band are the Fabry--P\'erot resonances of the finite
$N$-cell stack and are an unavoidable consequence of the finite
length of the composite region.

The crucial observation is that, at each fixed temperature, the
heating and cooling branches yield different transmission spectra.
At $T = 38~^\circ$C [panel (a)], the martensitic branch (blue)
displays its characteristic stop bands at the positions inherited
from the pure-phase Bloch structure of Fig.~\ref{fig:bandgapPhi}(a),
while the mixed-phase branch (purple) shows stop bands that are both
shifted to higher frequencies and noticeably wider, reflecting the
larger impedance contrast of the partially transformed alloy.
The reverse situation appears at $T = 48~^\circ$C [panel (b)],
where the mixed-phase branch (purple) and the fully austenitic
branch (red) again differ markedly. Across the frequency range
shown, there are intervals in which one branch transmits while the
other strongly attenuates, the magnitude of the difference
exceeding $50$~dB at several locations.

The role of the mixed-phase curve in linking the two panels is
worth highlighting. The purple curve appears in both
panels, in the cooling branch at $T = 38~^\circ$C and in the
heating branch at $T = 48~^\circ$C. These two curves describe the
same physical state ($\zeta = 0.5$) realized at temperatures that
differ by $10~^\circ$C, the width of the hysteresis loop. The
$10~^\circ$C separation between identical-spectrum lines in the two
panels is therefore a direct visual signature of the thermal
hysteresis of the underlying transformation.

The combination of Sec.~\ref{sec:infinite} and the present section
demonstrates that the hysteresis of the NiTiCu phase transition is
faithfully transferred to the acoustic response of the composite
rod, both in the intrinsic Bloch spectrum and in the experimentally
accessible transmission. The composite rod thus operates as a
finite phononic structure whose stop-band structure is selected by
thermal history, the central effect anticipated in Sec.~\ref{sec:material}.

\section{Conclusions}
\label{sec:conclusions}

We have shown that when one constituent of a phononic rod undergoes a first-order martensitic transformation, the periodic structure does more than tune with temperature: it converts the intrinsic thermal hysteresis of the material into a history-dependent acoustic response. A smooth, multivalued variation of the elastic modulus of NiTiCu is transformed by the periodic geometry into a sharp, history-selected transmission behavior. At a fixed probe frequency, a six-cell rod can switch between transmitting and strongly attenuating states, with contrasts exceeding 50 dB, depending solely on whether the sample was last heated or cooled. This behavior is reflected both in the Bloch spectrum of the infinite crystal and in the transmission spectrum of finite structures. The filling fraction provides an independent geometric control parameter, allowing the width of the hysteresis loops and the positions of gap closures to be tailored separately from temperature.

The model rests on a single phenomenological assumption: the heating branch of the temperature-dependent elastic modulus, E(T), is inferred rather than directly measured for the specific NiTiCu composition considered here. Although the existence of hysteretic band structures and history-dependent transmission does not depend on the precise width of the hysteresis loop, direct acoustic measurements of both thermal branches would remove this assumption and strengthen the quantitative predictions.

Several extensions suggest themselves naturally, including defect and graded structures, oblique incidence, and fully coupled thermoelastic descriptions. Likewise, corrections beyond the thin-rod approximation,  would modify the absolute positions of high-frequency gaps without altering the underlying hysteretic mechanism. The most immediate next step is experimental: immersion-ultrasonic transmission measurements on NiTiCu/Parylene C composite rods could provide a direct test of the predicted acoustic memory and hysteretic phononic band structures.
% =====================================================================
% Bibliography
% New references introduced in v2 are tagged "[NEW IN v2]".
% =====================================================================

\end{document}